\documentclass[12pt,preprint]{emulateapj}
\usepackage{graphics}

\begin{document}

\date{\today}
\title{Halo asphericity and the shear 3-pt function}
\author{Shirley Ho}
\author{Martin White}
\affiliation{Department of Physics, University of California, Berkeley,
CA 94720}

\begin{abstract}
We demonstrate through the use of a simple toy model that asphericity
in dark matter halos has a measurable effect on the configuration
dependence of the weak lensing shear 3-point function at small scales.
This sensitivity provides a way, in principle, to measure the shapes of
dark matter halos.  The distribution of halo ellipticities should be
included in models aiming at a high fidelity prediction of $n$-point shear
correlation statistics.
\end{abstract}

\maketitle

\section{Introduction}

Gravitational lensing (for reviews, see Mellier \cite{LensReview}; Bartelmann \& Schneider \cite{Schneider}; Wittman \cite{Wittman}) provides us with a unique
opportunity to probe the matter distribution of the universe and,
in combination with galaxy surveys, the galaxy-halo connection.
In particular the small angular scale structure of the lensing shear
field is sensitive to the density distribution of dark matter halos
over a range of mass scales, as has been emphasized by Takada \& Jain \cite{TakJai}.

In this brief report we show, using a simple toy model, that the 3-point
function of the shear is sensitive to the halo ellipticity/asphericity.
This gives us another way, in principle, to measure the shapes of dark
matter halos.
We expect that the higher order shear correlation functions will be ever
more sensitive to the asphericity of halos, thus the distribution of halo
shapes, in addition to the distribution of sizes, masses and radial profiles,
should be included in semi-analytic models aiming at a high fidelity
prediction of the higher order shear correlations.

\section{The halo model}

Our calculations will all be performed within the halo model paradigm
(see Corray \& Sheth \cite{HaloModel}).
Specifically we assume that all the mass in the universe is contained
within virialized halos with a range of sizes and shapes.
Under this assumption the correlation function and higher order moments
of the shear can be related to integrals over the (projected) density
profiles of the halos, times geometrical factors, and terms involving the
clustering of the halos (Takada \& Jain \cite{TakJaib}; Zaldarriaga \& Scoccimarro \cite{ZalSco}).
If we work on small (arcminute) angular scales then the dominant
contribution to the shear 3-point function will arise when all 3 points
sample the mass from within the same halo (Takada \& Jain \cite{TakJaib}).
This term is independent of the clustering of the halos.

\begin{figure}
\begin{center}
\resizebox{2.0in}{!}{\includegraphics{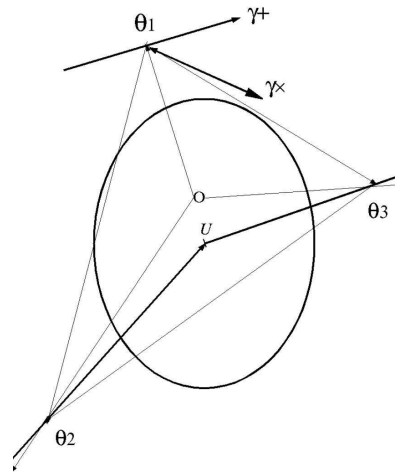}}
\end{center}
\caption{The configuration of the 3-point function and the (aspherical)
halo.  The vertices of the triangle are ${\bf \theta}_i$ with the center
at the origin.  The halo center is at ${\bf u}$.  We show the directions
of $\gamma_+$ and $\gamma_\times$ at ${\bf \theta}_1$.}
\label{fig:geom}
\end{figure}

To describe the 3-point function we will use conventions similar to those
by Zaldarriaga \& Scoccimarro \cite{ZalSco} and Schneider \& Lombardi \cite{Geomsource}, with ${\bf \theta}_i$ the vector to
vertex $i$ of the triangle with the origin of the coordinate system set to
the center of the triangle and  $\bf u$ is the center of the halo we are
considering. The geometry is shown in Fig.~\ref{fig:geom}.

We construct scalar $3$-point correlation functions by contracting the
three observed shears with spin-2 quantities (Mellier \cite{LensReview}; Bartelmann \& Schneider \cite{Schneider}; Wittman \cite{Wittman}; Zaldarriaga \& Scoccimarro \cite{ZalSco})
\begin{equation}
\begin{array}{lcccl}
 {\bf P}_+ &=& (\theta_x^2 - \theta_y^2 , 2\theta_x\theta_y )\theta^{-2}
           &=& (\phantom{-}\cos 2\phi, \sin 2\phi) \\
 {\bf P}_\times &=&
    (-2\theta_x\theta_y , \theta_x^2-\theta_y^2) \theta^{-2}
    &=& (-\sin 2\phi, \cos 2\phi)
\end{array}
\end{equation}
which have opposite parity.
When we contract the above quantities with the shear components $\gamma_1$
and $\gamma_2$, we will get $\gamma_+$ and $\gamma_\times$.
Thus, we can define our three point function as
\begin{equation}
  \zeta_{\alpha\beta\chi} =
  \left\langle \gamma_\alpha\gamma_\beta\gamma_\chi \right\rangle =
    {\bf P_\alpha^{\mu_1}} {\bf P_\beta^{\mu_2}} {\bf P_\chi^{\mu_3}}
  \left\langle \gamma_{\mu_1}\gamma_{\mu_2}\gamma_{\mu_3}\right\rangle
\end{equation}
where $\alpha$, $\beta$ and $\chi$ are either + or $\times$

The contribution to the small-angle $\zeta$ from a single halo of projected
density $\Sigma$ is then
\begin{equation}
  \zeta_{\alpha\beta\gamma} = \int d^2u\
   \epsilon^\alpha \Sigma({\bf\theta}_1-{\bf u})
   \epsilon^\beta  \Sigma({\bf\theta}_2-{\bf u})
   \epsilon^\gamma \Sigma({\bf\theta}_3-{\bf u})
\end{equation}
where the halo is centered at $\bf u$ and the $\epsilon^i$ are phase factors
that depends on the geometry of the halo.

It has been standard in the literature to use a ``spherically averaged''
halo profile in the computation of the higher-order moments of the shear
field.  In this limit the shear is tangential to the halo center and the
phase factors are simply
\begin{equation}
  \epsilon^+ = \cos 2 \alpha \qquad \epsilon^\times = \sin 2\alpha
  \qquad ,
\end{equation}
where $\alpha$ is the angle between ${\bf u}$ and ${\bf\theta}$.
However we know from numerical simulations
(Frenk \& White \cite{Frenk}; Dubinski \& Carlberg \cite{Dubin})
that dark matter halos are triaxial, so we wish to relax the
spherical assumption.

\begin{figure}
\begin{center}
\resizebox{3.5in}{!}{\includegraphics{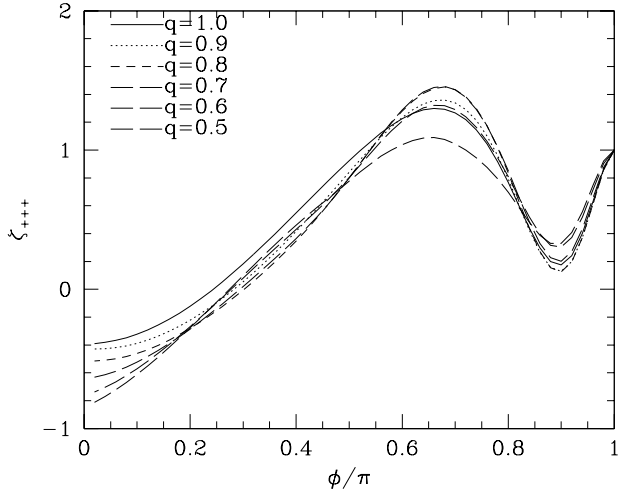}}
\end{center}
\caption{The 3-point correlation function, $\zeta_{+++}$ for an
isoceles triangle with opening angle $\phi$ for a series
of ellipticities $q$.  As the halo is made more elliptical, the peak
positions and heights change.  We have normalized all curves to unity
at $\phi=\pi$ for simplicity.}
\label{fig:3pt}
\end{figure}

In order to highlight the effects of halo asphericity on the 3-point function
with minimal complications we shall employ a simple toy model.
First, we work entirely in terms of a single halo, we do not integrate over
the entire mass distribution.  Second, we work throughout with projected
quantities.  Specifically we modify the spherical (projected) density profile,
$\Sigma(r)$, to $\Sigma(s)$ where $s=\sqrt{x'^2+(qy')^2}$ and $x'$ and
$y'$ are sky coordinates, possibly rotated with respect to the axes defining
the shear triangle.
For $q<1$ the profile is elliptical, elongated along the $x'$ axis.
Meneghetti, Bartelmann \& Moscardini \cite{Meneghetti} found $q\simeq 0.7$
characterized the ellipticity of dark matter halos found in N-body simulations,
and we shall use this value later as an indicator of the potential size of
ellipticity effects.
Since our profile is now elliptical, the phase factors, $\epsilon$, become
\begin{equation}
  \epsilon^+ = \cos 2(\alpha-\beta)
  \qquad \epsilon^\times = \sin 2(\alpha-\beta)
\end{equation}
where $\beta$ is the angle between $\theta-\bf u$ and the normal to the
tangent at $\theta$.
The definitions of the variables can be easily understood from
Figure \ref{fig:geom}.
Thus, for an NFW halo (Navarro, Frenk \& White \cite{NFW}),
our expression for $\zeta_{+++}$ becomes
\begin{equation}
  \zeta_{+++} = \int d\phi\int d^2u\
  \prod_{i=1}^3\cos 2(\alpha_i-\beta_i) \Sigma(s_i)
\end{equation}
where
\begin{eqnarray}
s_i &=& \sqrt{x_i^2+(qy_i^2)} \\
x_i &=&\phantom{-} (\theta_x^i-u_x)\cos\phi+(\theta_y^i-u_y)\sin\phi \\
y_i &=& -(\theta_x^i-u_x)\sin\phi+(\theta_y^i-u_y)\cos\phi ,
\end{eqnarray}
$\alpha_i$ and $\beta_i$ are functions of $\theta_i$ and $q$,
$\int d\phi$ is an integral over the orientation of the halo and
$\Sigma$ is the projected profile which we take to be (Bartelmann \cite{Bart})
\begin{equation}
  \Sigma^{2}(s) = \left(1\over s^2-1\right)
  \left(1-{2\over \vert s^2-1\vert^{1/2}}
  {\rm tann}^{-1}\left\vert {s-1\over s+1}\right\vert^{1/2}\right)
\end{equation}
where ${\rm tann}^{-1}(x)=\tan^{-1}(x)$ when $x>1$ and
$\tanh^{-1}(x)$ when $x<1$.
Similar expressions hold for the other components of $\zeta$.

\begin{figure}
\begin{center}
\resizebox{3.5in}{!}{\includegraphics{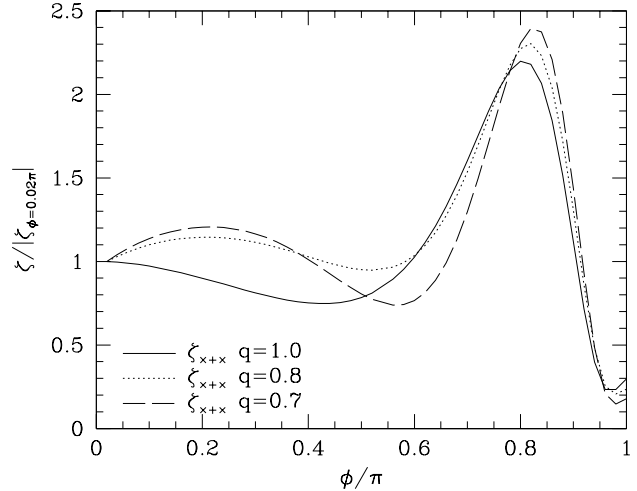}}
\end{center}
\caption{Another 3-point correlation function, $\zeta_{\times+\times}$,
for an isoceles triangle with opening angle $\phi$ for two
different ellipticities $q$.
As the halo is made more elliptical, the peak positions and heights change.
We have normalized all curves at $\phi=0.02\pi$ for simplicity.}
\label{fig:3pt2}
\end{figure}

\section{Results and Discussions}

It has been argued that halo ellipticity would not affect the higher
order correlations of the shear, since different orientations would
``wash out'' the effect. Our results show that this argument is too
simplistic -- the physical observables do depend on $q$.
One reason for this is the change in the projection of the shear at
$\theta_i$ onto the $+$ and $\times$ components.
The other is simply that the orientation average of $\rho^3$
isn't equal to the cube of the orientation averaged $\rho$, an inequality
which becomes stronger as we go to higher orders.

\begin{figure}
\begin{center}
\resizebox{3.5in}{!}{\includegraphics{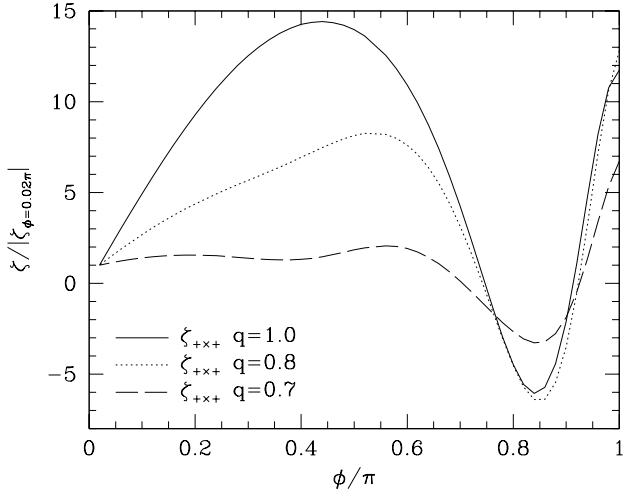}}
\end{center}
\caption{Another 3-point correlation function, $\zeta_{+\times+}$,
for an isoceles triangle with opening angle $\phi$ for two
different ellipticities $q$.
As the halo is made more elliptical, not only do the peak positions and
heights change, the shape of the function also changes.  We have normalized
all curves at $\phi=0.02\pi$ for simplicity.}
\label{fig:3pt3}
\end{figure}

Since $\zeta_{+++}$ has the strongest signal among all of the 3-point
functions, we have used it to illustrate the effect of halo asphericity.
The most noticeable property of Fig.~\ref{fig:3pt} is the peak shift and
amplitude change when $q<1$.  The peak first moves to the right from around
$2\pi/3$ and then moves to the left.  It may be possible to see such a
shift in Fig.~10 of Takada \& Jain \cite{TakJai} which compares the analytic
theory (assming spherical halos) to numerical results.
We can try to understand the shape of $\zeta_{+++}$ and the peak shift
by starting from a simple case when the halo is spherical and the triangle
is equilateral.
Let us consider the halo center to be inside the triangle, which minimizes
the distances to the vertices and thus maximizes the signal, and assume
that the triangle fits entirely within the halo.
For $q=1$ the configuration which best matches the shear orientations to
$\gamma_{+}$ at each vertix can easily be seen to be an equilateral triangle.
For a wide range of halo center positions one obtains a positive contribution
to $\zeta_{+++}$.
If the halo has $q<0.7$ then triangles with $\phi<2\pi/3$ have some
orientations which allow more of the high density region of the halo to
fall within the triangle and to better match the orientations of the $+$
shear components, enhancing the signal at slightly smaller angles than
the $q=1$ case.
For completeness we also show two of the seven other 3-point functions in
Figs.~\ref{fig:3pt2} and \ref{fig:3pt3}.
To avoid clutter we only show results for $q=1$, $0.8$ and $0.7$.
These figures show clearly that halo asphericity has an effect on all of
the (non-zero) shear 3-point functions.

We have shown in this brief report that aspherical dark matter halos
give rise to a different configuration dependence of the weak lensing
shear 3-point function than spherical halos.  The positions and amplitudes
of the peaks in the 3-point function shift as the typical halo is made more
aspherical.  This effect, in principle, provides us a probe of dark matter
halo shapes, which have long been predicted by numerical simulations to be
aspherical.  It should also be included in semi-analytic models aiming at
a high fidelity prediction of higher order correlations in the shear field.

\begin{acknowledgments}
SH wants to thank Joanne Cohn, Chuck Keeton, Chris Vale, Amol Upadhye and
the Berkeley Cosmology Group for their valuable discussions and support.
This research was supported by the NSF and NASA.
\end{acknowledgments}

\end{document}